\documentclass{article}
\usepackage{amsmath,mathtools}

\newcommand{\xddots}{%
  \raise 4pt \hbox {.}
  \mkern 6mu
  \raise 1pt \hbox {.}
  \mkern 6mu
  \raise -2pt \hbox {.}
}

\usepackage{alltt}
\usepackage[vcentermath]{youngtab}
\usepackage[all]{xy}
\usepackage{latexsym,amssymb,amsmath,amsfonts, amsthm, xcolor}

\def\vbar{\mathchoice{\vrule height6.3ptdepth-.5ptwidth.8pt\kern-.8pt}
  {\vrule height6.3ptdepth-.5ptwidth.8pt\kern-.8pt}
  {\vrule height4.1ptdepth-.35ptwidth.6pt\kern-.6pt}
  {\vrule height3.1ptdepth-.25ptwidth.5pt\kern-.5pt}}
\def\fudge{\mathchoice{}{}{\mkern.5mu}{\mkern.8mu}}
\def\bbc#1#2{{\rm \mkern#2mu\vbar\mkern-#2mu#1}}
\def\bbb#1{{\rm I\mkern-3.5mu #1}}
\def\bba#1#2{{\rm #1\mkern-#2mu\fudge #1}}
\def\bb#1{{\count4=`#1 \advance\count4by-64 \ifcase\count4\or\bba A{11.5}\or
  \bbb B\or\bbc C{5}\or\bbb D\or\bbb E\or\bbb F \or\bbc G{5}\or\bbb H\or
  \bbb I\or\bbc J{3}\or\bbb K\or\bbb L \or\bbb M\or\bbb N\or\bbc O{5} \or
  \bbb P\or\bbc Q{5}\or\bbb R\or\bbc S{4.2}\or\bba T{10.5}\or\bbc U{5}\or
  \bba V{12}\or\bba W{16.5}\or\bba X{11}\or\bba Y{11.7}\or\bba Z{7.5}\fi}}

\newcommand{\vs}{\vspace{0.25cm}}

\newtheorem{theorem}{Theorem}
\newtheorem{itlemma}{Lemma}[section]
\newtheorem{itproposition}[itlemma]{Proposition}
\newtheorem{itcorollary}[itlemma]{Corollary}
\newtheorem{itremark}[itlemma]{Remark}
\newtheorem{itremarks}[itlemma]{Remarks}
\newtheorem{itdefinition}[itlemma]{Definition}
\newtheorem{itexample}[itlemma]{Example}

\newenvironment{lemma}{\begin{itlemma}\rm}{\end{itlemma}} 
\newenvironment{remark}{\begin{itremark}\rm}{\end{itremark}} 
\newenvironment{remarks}{\begin{itremarks} \rm}{\end{itremarks}}
\newenvironment{corollary}{\begin{itcorollary}\rm}{\end{itcorollary}}
\newenvironment{proposition}{\begin{itproposition}\rm}{\end{itproposition}}
\newenvironment{definition}{\begin{itdefinition}\rm}{\end{itdefinition}}
\newenvironment{example}{\begin{itexample}\rm}{\end{itexample}}
\newenvironment{fact}{\noindent {{\bf Fact}}:\ \ }{\hfill \medskip}
\newenvironment{claim}{\noindent {\em Claim}. \ \ }{\hfill \medskip}
\newcommand{\be}[1]{\begin{equation}\label{#1}}
\newcommand{\ee}{\end{equation}}
\newcommand{\bl}[1]{\begin{lemma}\label{#1}}
\newcommand{\br}[1]{\begin{remark}\label{#1}}
\newcommand{\brs}[1]{\begin{remarks}\label{#1}}
\newcommand{\bt}[1]{\begin{theorem}\label{#1}}
\newcommand{\bd}[1]{\begin{definition}\label{#1}}
\newcommand{\bp}[1]{\begin{proposition}\label{#1}}
\newcommand{\bc}[1]{\begin{corollary}\label{#1}}
\newcommand{\bfact}[1]{\begin{fact}\label{#1}}
\newcommand{\bex}[1]{\begin{example}\label{#1}}
\newcommand{\ec}{\end{corollary}}
\newcommand{\efact}{\end{fact}}
\newcommand{\eex}{\end{example}}
\newcommand{\el}{\end{lemma}}
\newcommand{\er}{\end{remark}}
\newcommand{\ers}{\end{remarks}}
\newcommand{\et}{\end{theorem}}
\newcommand{\ed}{\end{definition}}
\newcommand{\ep}{\end{proposition}}
\newcommand{\epr}{\end{proof}}
\newcommand{\bpr}{\begin{proof}}
\newcommand{\bcl}{\begin{claim}}
\newcommand{\ecl}{\end{claim}}

\newcommand{\bi}{\begin{itemize}}
\newcommand{\ei}{\end{itemize}}
\newcommand{\ben}{\begin{enumerate}}
\newcommand{\een}{\end{enumerate}}

\usepackage{graphicx}

\title{SUBSPACE CONTROLLABILITY OF MULTI-PARTITE SPIN NETWORKS}
\author{Francesca Albertini\thanks{Dipartimento di Tecnica e Gestione del Sistemi Industriali, Universita' di Padova, Vicenza, Italy, albertin@math.unipd.it}       $\qquad$ and  $\qquad$ Domenico D'Alessandro\thanks{Department of Mathematics, Iowa State University, daless@iastate.edu} }
\date{\today}

\begin{document}

\maketitle 
\begin{abstract} 

In a network of spin $\frac{1}{2}$ particles, controlled through an external electro-magnetic field, the  gyromagnetic ratio of each spin is a parameter that characterizes the interaction of the spin  with the external control field. {\it Multipartite} networks are such that the spins are divided into subsets according to their gyromagnetic ratio and spins in one set interact in the same way with all spins in another set. Due to the presence of symmetries in this type of systems,  the underlying Hilbert state space splits into invariant subspaces for the dynamics. Subspace controllability is verified if every unitary evolution can be generated by the dynamics on these subspaces.  

We give an exact characterization, in term of graph theoretic conditions,  of subspace controllability for multipartite quantum spin networks. This extends and unifies  previous results.

\end{abstract}

\vs 

\vs
\vs

\vs 

\vs
\vs

\section{Introduction and statement of main result}

The dynamics of quantum mechanical systems, subject to a control electromagnetic field, can often be described by  the Schr\"odinger equation in the form
\be{Scro1}
\dot \psi=A \psi+\sum_{j=1}^mB_j u_j \psi,  
\ee
where $u_j$, $j=1,...,m$, are the control variables and $\{A,B_1,...,B_m\}$ are given operators,  with $\psi$ denoting the state of the quantum system, varying in the underlying Hilbert space ${\cal H}$. In finite dimensions, the controllability properties of system (\ref{Scro1}) are usually assessed using the {\it Lie algebra rank condition} (see, e.g., , \cite{HT}, \cite{JS}). One calculates the Lie algebra ,  ${\cal G}$,  generated by the matrices $\{A,B_1,...,B_m\}$, which is called the {\bf dynamical Lie algebra}.  Given $e^{\cal G}$ the connected  Lie group associated with it, assumed compact, the condition says that the reachable set ${\cal R}_{\psi_0}$, for (\ref{Scro1}) starting from $\psi_0$ is given by 
$$
{\cal R}_{\psi_0}=\{ X \psi_0 \, | \, X \in e^{\cal G}\}. 
$$
In the case of large systems,  it is important to find ways to assess controllability which avoid the repeated calculation of commutators of very large matrices in (\ref{Scro1}). Such controllability criteria should be 
 easily related  to  the physical structure of the system under consideration. One example of large system is given by  networks of $n$ interacting spin $\frac{1}{2}$ particles, where the dimension of the Hilbert space ${\cal H}$  grows exponentially with $n$,  as $2^n$. In some cases, {\it graph theoretic conditions} have been given to assess the controllability of quantum systems (see, e.g., \cite{NoiLAAold}, \cite{Turinici}), and this paper has this objective as well.

In the presence of a group of symmetries $G$, i.e., a (discrete) group of matrices commuting with the matrices $\{A,B_1,...,B_m\}$ in (\ref{Scro1}), the underlying Hilbert space ${\cal H}$ for the system  
splits in the direct sum of invariant subspaces for the dynamics (\ref{Scro1}) and, in an appropriate basis, the matrices $\{A,B_1,...,B_m\}$ in (\ref{Scro1}) take a block diagonal form \cite{ConJonas}. Transitions from one subspace to the other are forbidden and therefore controllability is lost. For a network of $n$ spins, the topology of the network  itself often suggests the symmetries to be considered which typically are subgroups of the permutation groups leaving the 
network unchanged. For example, for the network of Figure \ref{F1} the group of permutations on the three  spin 
in the set $Cl_3$ is a symmetry group for the system. In these cases, it is of interest to investigate whether one has controllability within the invariant subspaces. This property is called {\bf subspace controllability} and it has been investigated in several recent papers (see, e.g., \cite{NoiLAA}, \cite{Xinhua1}, \cite{Dan1}, \cite{Dan2} ). We shall see  that, in general,  this property is subspace-dependent, that is, for the same decomposition,  there might be some subspaces of dimension $D$ where the restriction of the dynamical Lie algebra is the {\it full} $su(D)$ (special unitary) Lie algebra of dimension $D$ and some others where it is not. In the first case subspace controllability is verified, in the second case it is not.

In this paper, we shall explore subspace controllability for networks of spin $\frac{1}{2}$ particles in the multipartite configuration. This means that spin particles are collected in sets, which we shall call {\it  clusters} according to the value of their {\it gyromagnetic ratio}, that is the parameter which models the interaction with an external control field.  Spins in the same cluster  interact in the same way with spins in another cluster. These systems presents a group of symmetries given  by permutations of the spins within the same cluster. We shall give in Theorem \ref{MainT} a necessary and sufficient condition of subspace controllability for such systems in graph theoretic terms. This result will extend the result of \cite{NoiLAA} which only dealt with the bipartite case and with bounds on the number of spin in one cluster. We remove such bounds. The technique we shall use is different from the one in \cite{NoiLAA} which was based on a direct computation of the dynamical Lie algebra. Here we shall use techniques of representation theory and, in particular, the{\it  Clebsch-Gordan decomposition}  (see, e.g.,\cite{FH},  \cite{Ramond},  \cite{Woit}) of the tensor product representation of $su(2)$. Our result also generalizes  the result of \cite{NoiLAAold} which is found as a special case when all the spins have different gyromagnetic ratios.

\subsection{Basic notations}

{

We recall the definition of the Pauli matrices 

\be{PauliMat}
\sigma_x:=\frac{1}{2} \begin{pmatrix} 0 & 1 \cr 1 & 0  \end{pmatrix}, \qquad \sigma_y:= 
\frac{1}{2}
 \begin{pmatrix} 0 & i \cr -i & 0  \end{pmatrix}, \qquad
 \sigma_z:= \frac{1}{2}  \begin{pmatrix}  1 & 0 \cr 0 & -1\end{pmatrix}, 
\ee
which satisfy the basic commutation relations 

\be{commurel}
[i\sigma_x,i\sigma_y]=i\sigma_z,  \qquad  [i\sigma_y,i\sigma_z]=i\sigma_x, \qquad [i\sigma_z,i\sigma_x]=i\sigma_x, 
\ee
and 
\be{basic2}
\sigma_x^2=\sigma_y^2=\sigma_z^2=\frac{1}{4}{\bf 1}, \qquad \{ \sigma_x ,\sigma_y\}=\{ \sigma_y, \sigma_z\}=\{ \sigma_z,  \sigma_x\}=0,    
\ee
where $\{A, B\}$ is the {\it anticommutator} of $A$ and $B$, that is,  $\{A,B\}:=AB+BA$. 
Here and in the following ${\bf 1}$ always denotes the identity matrix or operator, the dimension being understood from the context. The matrices $\{i\sigma_x, i\sigma_y, i\sigma_z\}$ along with the commutation relations (\ref{commurel}) form an irreducible 2-dimensional representation of $su(2)$,  the {\it standard representation}. 
Given a certain positive integer $\tilde{n}$, which is usually determined by the context, 
the matrices $S_{x,y,z}$ are defined as the sums of $\tilde{n}$ terms  where each term 
is the tensor product of $\tilde n$ factors,  each being the $2 \times 2$ identity except the $l$-th one which is  $\sigma_{x,y,z}$, for $l=1,2,...,\tilde{n}$. So, for example, for $\tilde{n}=3$, 
$$
S_x=\sigma_x \otimes {\bf 1} \otimes {\bf 1}+ {\bf 1} \otimes \sigma_x \otimes {\bf 1} +
{\bf 1} \otimes  {\bf 1} \otimes {\sigma_x}. 
$$
We denote by $I_{gb}$ the sum of matrices where each term of the sum is the tensor product of  $\tilde{n}$ identities 
except for one position 
occupied by $\sigma_g$ and one occupied by $\sigma_b$ and viceversa. The sum extends over all possible pairs of locations and therefore contains $\tilde{n}(\tilde{n}-1)$ terms. For example for $\tilde{n}=3$, $I_{xy}$ is equal to 
$$
I_{xy}=\sigma_x \otimes \sigma_y \otimes {\bf 1}+ \sigma_y\otimes \sigma_x \otimes {\bf 1}+
\sigma_x \otimes {\bf 1} \otimes \sigma_y + \sigma_y\otimes {\bf 1} \otimes \sigma_x +
{\bf 1} \otimes \sigma_x \otimes \sigma_y + {\bf 1} \otimes \sigma_y \otimes \sigma_x 
$$

{
We consider a network of  spin $\frac{1}{2}$ particles grouped in  
$N$ {\bf clusters} of indistinguishable  spins. Clusters are defined as sets of spin particles which have the same gyromagnetic ratios.
Moreover, we assume that each spin in a cluster interacts in the same way with spins in a different cluster and do not interact with each other. Any permutation of the spins belonging to the same cluster will leave the dynamics unchanged.
}  

If a network has $N$ clusters, with the $k$-cluster having $n_k$ spin particles,  we denote 
by $A^{j}$ a  matrix 
which is the tensor product of $N$  identity matrices, where in the $k$-cluster the identity has  dimension $2^{n_k}$, except   in the position $j$ which is occupied by the matrix $A$, a $2^{n_j} \times 2^{n_j}$ matrix.
 Examples of these types of matrices we shall often use are $S_{x,y,z}^j$, { $j=1,\ldots,N$. So, for example:
 \[S^2_x={\bf 1} \otimes  S_x\otimes {\bf 1} \otimes\cdots \otimes  {\bf 1},\]
where $S_x$ has dimension  $2^{n_2}$  and the identity matrix in the first position has dimension $2^{n_1}$, the one in the third position has dimension $2^{n_3}$, and so on.  }
 
 Extending this notation,  the matrices of the form $A^{j} B^k$, with $j\not=k$, $j,k \in \{1,2,...,N\}$, can be seen as the  product of $A^j$ and $B^k$ but also as tensor products of identities, with various dimensions, except in the positions $j$ and $k$ occupied by $A$ and $B$, respectively, of dimensions $2^{n_j}$ and $2^{n_k}$. This notation is naturally  extended to any number of factors in the product besides two.

\subsection{The model}\label{TM}

The quantum control system model we shall study in this paper is a network of spin $\frac{1}{2}$ particles interacting with each other. We have grouped the spins in $N$ clusters of indistinguishable spins, each interacting with the same coupling constant with spins in other clusters. The interaction is assumed to be of the Ising $Z-Z$ form {($S^j_z S^k_z$) (although the results will be extended to every other type of two body interaction (coupling) in section \ref{Exte})}. The network is represented by a {\bf connectivity graph} where each node represents a cluster  of equivalent spins and there is an edge connecting two nodes if there is a non zero interaction between spins in the corresponding clusters. We assume the interactions between spin in two different clusters all equal.   For example, the network of Figure  \ref{F1}  consists of a total of eight spin $\frac{1}{2}$ particles, two of them in the first cluster ($Cl_1$), two of them in the second one ($Cl_2$), three of them in the third one ($Cl_3$) and one in the fourth one ($Cl_4$). The lines represent nonzero interactions which are assumed to be the same for spins belonging to the same couple of clusters. The connectivity graph for such a network is given in Figure \ref{F2}.

\begin{figure}[h]
\centerline{\includegraphics[width=4.8in,height=2.5in]{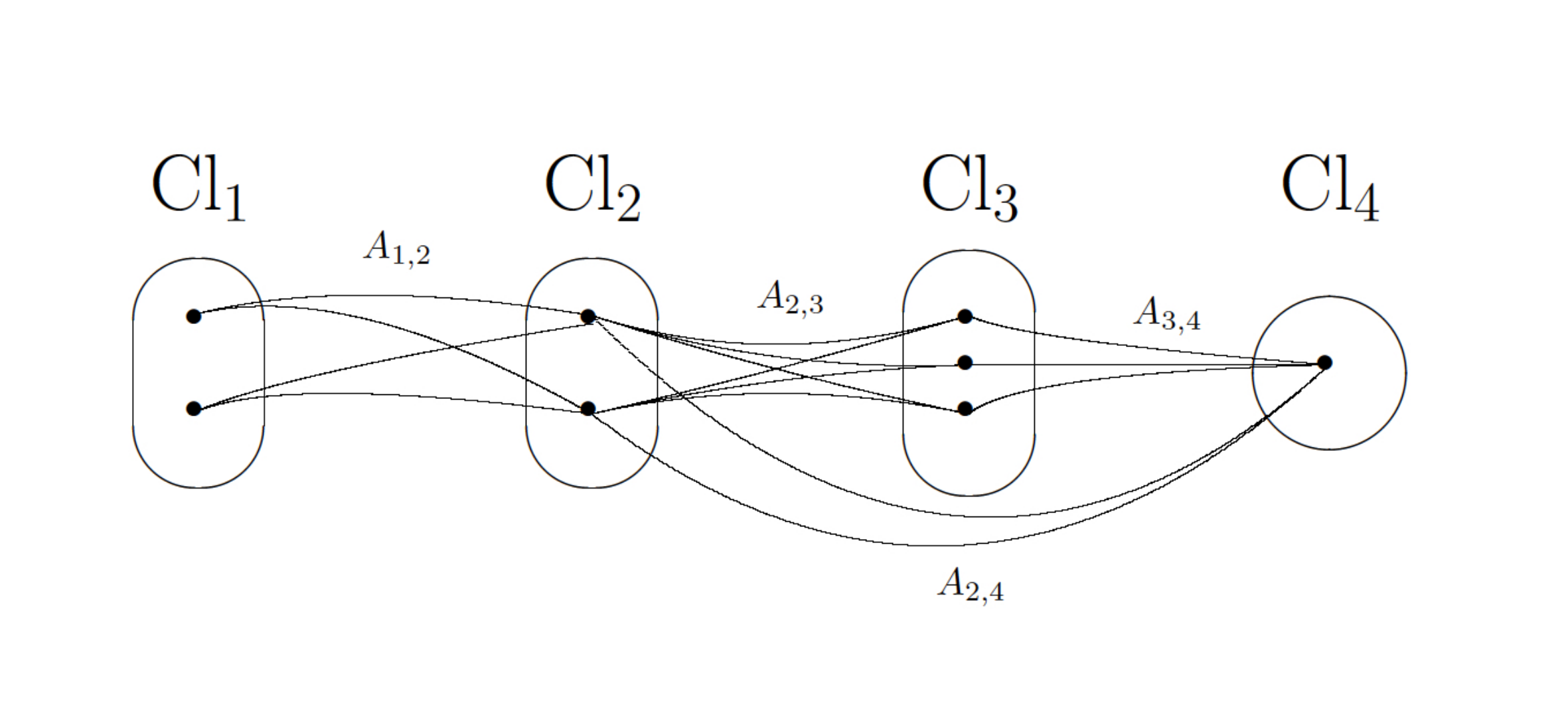}}
\caption{Example of a multipartite spin network}
\label{F1}
\end{figure}
\begin{figure}[h]
\centerline{\includegraphics[width=4.8in,height=2.5in]{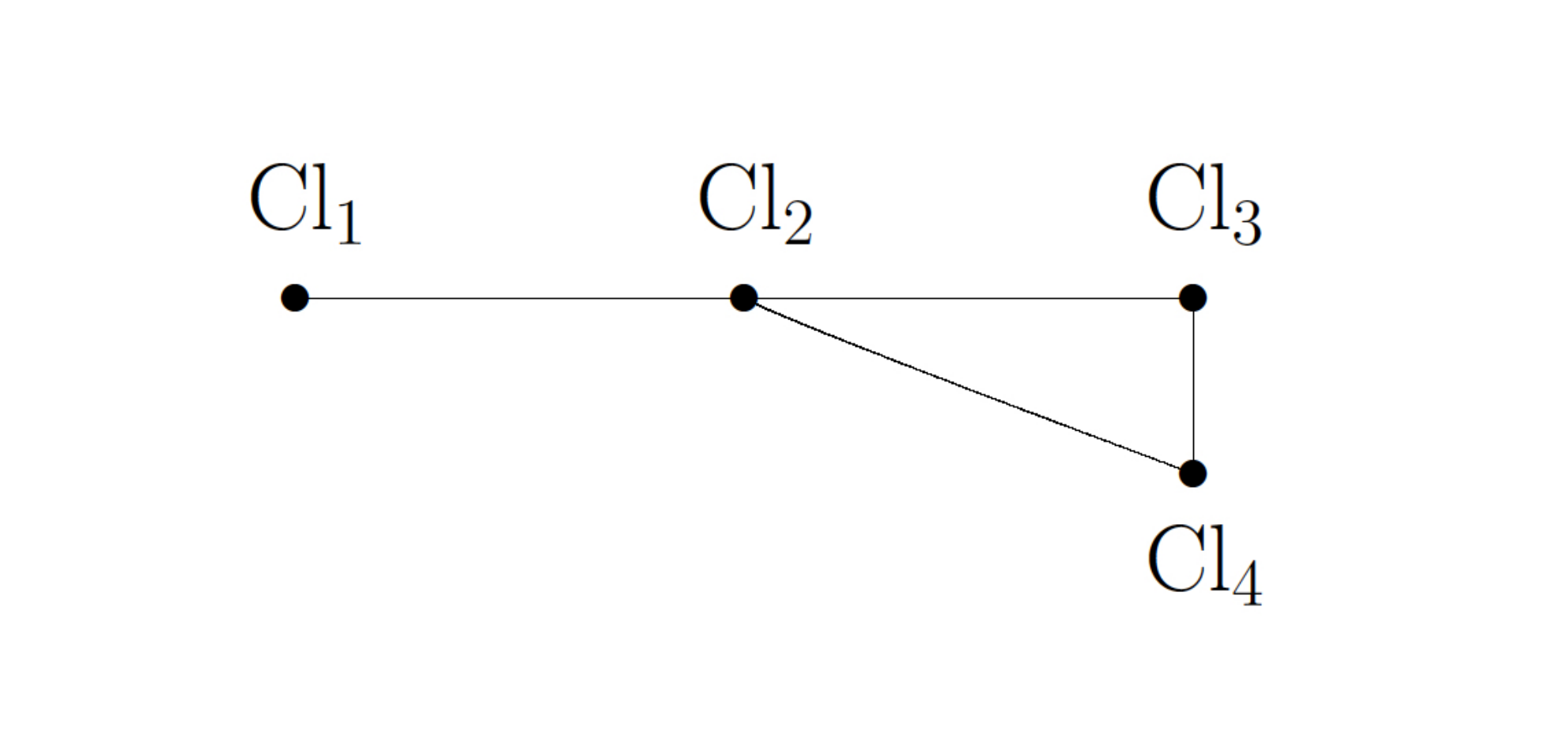}}
\caption{Connectivity graph for the network of Figure \ref{F1}}
\label{F2}
\end{figure}
The Schr\"odinger equation which models the dynamics takes the form (\ref{Scro1}) 
with 
\be{A}
iA:=\sum_{1\leq j < k \leq N} A_{j,k}S_z^jS_z^k, 
\ee
with $A_{j,k}$ the {\it coupling constants}  and 
\be{B}
iB_{x,y,z}=\sum_{j=1}^N\gamma_j S_{x,y,z}^j, 
\ee
where $\gamma_j$ are the {\it gyromagnetic ratios} of the spins in the cluster $j$, assuming an isotropic type of interaction with the three components of the electro-magnetic field $u_{x,y,z}$. 

We assume that some of the coupling constants $A_{j,k}$ are different from zero so that the {\em {connectivity graph associated with the network is connected.}} This is done without loss of generality since if the graph has several connected components we can repeated the analysis we shall perform on each one of them.

The dynamical Lie algebra ${\cal G}$, for this type of systems,
 is generated by $\{A,B_{x}, B_y, B_z \}$, in (\ref{A}) and (\ref{B}). 
 
 A crucial observation  for our development is that, with ${n}$ spin particles, 
  $\{ i S_x, iS_y,\, {\text{and}}\, iS_z\}$ span a $2^{{n}}$-dimensional representation of $su(2)$ since they satisfy 
  (cf. (\ref{commurel}))
  
 \be{commurel2}
[iS_x,iS_y]=iS_z,  \qquad  [iS_y,iS_z]=iS_x, \qquad [iS_z,iS_x]=iS_x. 
\ee
This representation coincides with the tensor product of ${n}$ copies of the standard representations (see, e.g., \cite{Woit}) as it will be further elaborated upon in the following. 

By using (\ref{commurel2}), we have that
\[
[B_{x},B_{y}]=\sum_{j=1}^N\gamma^2_j S_{z}^j, \quad [B_{y},B_{z}]=\sum_{j=1}^N\gamma^2_j S_{x}^j, \quad  [B_{z},B_{x}]=\sum_{j=1}^N\gamma^2_j S_{y}^j,
\]
belong to ${\cal G}$, and by iterating the Lie brackets, we have that all 
\[
\sum_{j=1}^N\gamma^l_j S_{z}^j, \qquad \sum_{j=1}^N\gamma^l_j S_{x}^j, \qquad  \sum_{j=1}^N\gamma^l_j S_{y}^j,
\]
for $l \geq 1$, belong to ${\cal G}$. Using a Vandermonde determinant type of argument and assuming, as we will,  that 
the $\gamma_{j}$'s are all different from zero (besides being different from each other), it follows that $iS_{x,y,z}^j$ for $j=1,...,N$,  also belong to ${\cal G}$.  Therefore, the dynamical  
Lie algebra ${\cal G}$ is generated by 
$$
{\cal S}:=\{ iS_x^j,i S_y^j, iS_z^j |\,  j=1,...,N\}, \qquad A=-i \sum_{1\leq j < k \leq N} A_{j,k}S_z^jS_z^k.
$$
We also have 

\bl{Lemma1}
The Lie algebra ${\cal G}$ is the same as the one generated by ${\cal S}$ and by all the 
$ i S^j_z S_z^k $ such that $A_{j,k}\not=0$.  
\el
\bpr
Set $j=1$ and $k=2$, without loss of generality and assume $A_{1,2}\not=0$. We want to show that $i S^1_zS^2_z$ belongs to ${\cal G}$. Start with  $[A, i S^1_x]$ to obtain $H_1:=-i \sum_{l>1}A_{1l} S_y^1 S_z^l$. Then take $[H_1, i S^1_x]$ to obtain $H_2=i \sum_{l>1}A_{1l} S_z^1 S_z^l$. Then take $[H_2, i S_x^2]$ to obtain $H_3=-iA_{12} S_z^1 S_y^2$. Then take $[H_3,i S_x^2]$ to obtain $iA_{12} S_z^1 S_z^2$. Since $A_{12}\not=0$, we obtain the result. 
\epr

\subsection{Decomposition in invariant subspaces and  subspace controllability}

Let $n_j$ denote the number of spins in the $j$-th cluster.  According to the postulates of quantum mechanics the subsystem corresponding to the $j$-th cluster lives in a Hilbert space $(V^1)^{\otimes n_j}$ where $V^1$ denotes the two dimensional (spin $\frac{1}{2}$) carrier of the standard representation of $su(2)$. The full space Hilbert state space is therefore 
 
\be{fullspace}
{\cal H}=(V^1)^{\otimes n_1} \otimes  (V^1)^{\otimes n_2} \otimes \cdots \otimes (V^1)^{\otimes n_N}.
\ee
Extending the above notation, let us  denote  by $V^l$ the spin $\frac{l}{2}$ irreducible representation of $su(2)$.  Here $V^l$ has (complex) dimension $l+1$. 

Using (iteratively) {\bf Clebsch-Gordan decomposition} (see, e.g., \cite{Woit})  we have that $(V^1)^{\otimes n_j}$  decomposes in the direct sum of a number of (possibly repeated) subspaces $V^{n_j}$, $V^{n_j-2}$,..., where the last term is $V^0$ or $V^1$ according to 
whether $n_j$ is even or odd, respectively. It is not important for our purposes how many copies of the same $V^l$ are present. This will be determined on a case by case basis according to the iteration for the given cluster. { For a fixed cluster $j$, the matrices 
$S^j_{x,y,z}$ act on each space $V^l$ as the $\frac{l}{2}$ irreducible representation of $su(2)$. 
In particular  when $l=0$ they have value equal to zero. This will be used in the following.}

\bex{fromeex}
Consider the network of spins of Figure \ref{F1} and the first cluster for which Clebsch-Gordan decomposition gives $V^1\otimes V^1=V^{1+1}\oplus V^{1+1-2}=V^2 \oplus V^0$. For the third cluster Clebsch-Gordan decomposition gives
$$
V^1 \otimes V^1 \otimes V^1=(V^2 \oplus V^0) \otimes V^1=(V^2 \otimes V^1) \oplus (V^0 \oplus V^1)=
V^3\oplus V^1 \oplus V^1.
$$
For the second cluster, we have $V^1 \otimes V^1=V^2\oplus V^0$ and for the fourth cluster, we have $V^1$. 
\eex 

We consider as {\em{ invariant }} subspaces of the full  system 
of $N$ clusters of spins  the spaces 
\be{subspacesF}
S =F_1 \otimes F_2 \otimes \cdots \otimes F_N, 
\ee
where $F_j$, $j=1,..,N$, is one of the spaces  $V^{n_j}$, $V^{n_j-2}$,.... The spaces (\ref{subspacesF}) are indeed invariant under the dynamical Lie algebra ${\cal G}$ since they are invariant under the generators. {We shall see later (see Remark \ref{minimality}) that they are {\it minimal}  invariant, that is,  they contain no proper nontrivial invariant subspaces. In the language of representation theory, they carry {\it irreducible}   representations of the dynamical Lie algebra ${\cal G}$.}

{\color{black}  As a result of the Clebsch-Gordan decomposition on each factor corresponding to a cluster the full Hilbert space ${\cal H}$ in (\ref{fullspace}) decomposes into the direct sum of invariant spaces of the form (\ref{subspacesF}). We can then take a basis of the full Hilbert space ${\cal H}$  by putting together the (orthogonal) bases of the subspaces of the type (\ref{subspacesF})}. In this  basis 
the dynamical Lie algebra ${\cal G}$ takes a block diagonal form. 

The dimension of each subspace $S$ in (\ref{subspacesF})  is 
\be{dimensioni}
D^S:=\dim(F_1)\dim(F_2)\cdots \dim(F_N). 
\ee
{ Subspace controllability} is a feature of each invariant subspace in (\ref{subspacesF}).
\bd{SCdef}
An invariant subspace (\ref{subspacesF}) is said to be {\bf subspace controllable}  if and only if,   for every $M$ in $su(D^S)$,  there exists a matrix in ${\cal G}$ such that its restriction to $S=F_1 \otimes F_2 \otimes \cdots \otimes F_N$ in  (\ref{subspacesF})  is equal to $M$. The full system is called subspace controllable if every invariant subspace is subspace controllable. More generally we define a {\bf subspace dynamical Lie algebra} ${\cal G}_S$ for the subspace (\ref{subspacesF}) as the largest Lie subalgebra of $su(D^S)$ such that for every matrix $M \in {\cal G}_S$ there exists an element in ${\cal G}$ whose restriction to $S=F_1 \otimes F_2 \otimes \cdots \otimes F_N$ is equal to $M$. Subspace controllability is verified when ${\cal G}_S=su(D^S)$. 
\ed

\subsection{Statement of main result}

The subspace dynamical Lie algebra, and therefore subspace controllability, can be assessed using a graph associated with the invariant subspace (\ref{subspacesF}) which we shall call the {\bf associated graph}. Such a graph is obtained from the connectivity graph of the spin network by removing the nodes corresponding to values of $j$ such that $F_j=V^0$ in (\ref{subspacesF}) and all the edges having such nodes as endpoint. Even if the original connectivity graph was connected (as we have assumed) the resulting  associated graph for  a subspace   (\ref{subspacesF}) might not be be connected, and, in general, it will have a number $m_c$ of connected components 
${\cal C}_1$, ${\cal C}_2$,...,${\cal C}_{m_c}$. We define the dimension associated with $h$-th  connected component,  as (cf., (\ref{dimensioni}))
\be{dimensioni2}
D^S_h:=\prod_{j \in {\cal C}_h} \dim(F_j). 
\ee
In the special case where $m_c=1$, we have only $D^S_1$ which coincides with $D^S$ in (\ref{dimensioni}).

\bex{fromeex2} Reconsider the network of Example \ref{fromeex} and Figures \ref{F1} \ref{F2} for which we have calculated the decompositions  for any cluster as 
$V^2 \oplus V^0,$  $V^2 \oplus V^0,$ $V^3 \oplus V^1 \oplus V^1, $ $V^1. 
$
The possible invariant subspaces (\ref{subspacesF})  are $T_{2,2,3,1}:=V^2\otimes V^2 \otimes V^3 \otimes V^1$,  $T_{2,2,1,1}:=V^2\otimes V^2 \otimes V^1 \otimes V^1$,  $T_{2,0,3,1}:=V^2\otimes V^0 \otimes V^3 \otimes V^1$, $T_{2,0,1,1}:=V^2\otimes V^0 \otimes V^1\otimes V^1$, $T_{0,2,3,1}:=V^0\otimes V^2 \otimes V^3 \otimes V^1$, $T_{0,2,3,1}:=V^0\otimes V^2 \otimes V^3 \otimes V^1$,  $T_{0,2,1,1}:=V^0\otimes V^2 \otimes V^1 \otimes V^1$, $T_{0,0,3,1}:=V^0\otimes V^0 \otimes V^3 \otimes V^1$,  $T_{0,0,1,1}:=V^0\otimes V^0 \otimes V^1 \otimes V^1$. In Figure \ref{Fig2} we report the associated graphs for $T_{2,2,3,1}$ (which coincides with the connectivity graph), $T_{2,0,3,1}$, and $T_{0,2,1,1}$. 
\eex

\begin{figure}[h]
\centerline{\includegraphics[width=4.8in,height=3.5in]{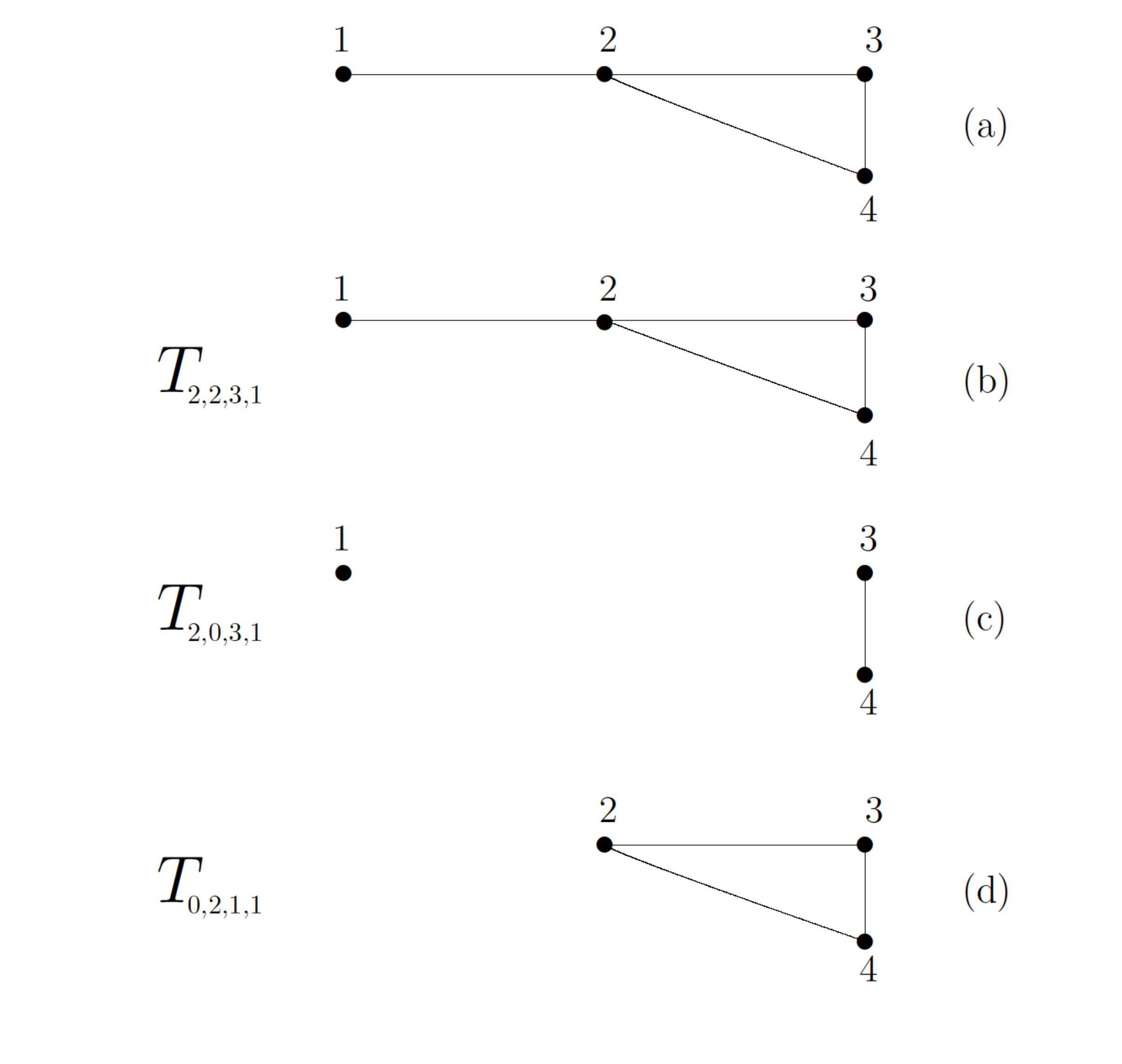}}
\caption{Associated graphs for invariant subaspaces $T_{2,2,3,1}$ (b), $T_{2,0,3,1}$ (c),  
$T_{0,2,1,1}$ (d), as compared with the connectivity graph of the network in part (a). }
\label{Fig2}
\end{figure}

\vspace{0.25cm}

The following result is the main theorem of this paper. It  allows to  characterize the subspace dynamical Lie algebra and therefore subspace controllability in every case.

\bt{MainT}
Consider an invariant subspace of the form (\ref{subspacesF}) and its associated graph with $m_c$ connected components. Then,  the subspace dynamical Lie algebra 
${\cal G}_S$ has the form of a direct sum 
\be{formaSDLA}
{\cal G}_S={\cal G}_1 \otimes {\bf 1} \otimes \cdots \otimes {\bf 1}+{\bf 1} \otimes 
{\cal G}_2 \otimes {\bf 1} \otimes \cdots \otimes {\bf 1}+\cdots + 
{\bf 1} \otimes \cdots \otimes {\bf 1} \otimes {\cal G}_{m_c}, 
\ee
where ${\cal G}_h$ is a Lie algebra acting on the space given by 
$
{\large{\otimes}}_{j \in {\cal C}_h} F_j. 
$
This space has 
dimension $D^S_h$ in (\ref{dimensioni2}) and it corresponds 
to the $h$-th connected component in the associated 
graph to (\ref{subspacesF}). 
In (\ref{formaSDLA}) ${\cal G}_h$, $h=1,...,m_c$  is (modulo multiples of the identity) 

\begin{enumerate}

\item Equal to the $\dim(F_j)$-irreducible representation of $su(2)$ if ${\cal C}_h$ only contains one node, the node $j$.

\item Equal to $su(D^S_h)$ if  ${\cal C}_h$ contains more than one node. 

\end{enumerate} 

\et

From the above theorem the following exact characterization of subspace controllability follows. 

\bc{exact}
A subspace (\ref{subspacesF}) is subspace controllable if and only if the associated graph is connected and contains at least two nodes. 
\ec

\bex{continuedz} 
Consider the subspaces of Example \ref{fromeex2}   with the 
associated graphs reported in Figure \ref{Fig2}. According to Corollary \ref{exact} subspace controllability is verified 
in the cases $T_{2,2,3,1}$ and $T_{0,2,1,1}$. It is not verified in the case of $T_{2,0,3,1}$. In this case,  on the given subspace, the subspace dynamical Lie algebra is the direct sum of two subalgebras, one subalgebra given by the irreducible representation of $su(2)$ on $V^2$, i.e., a representation of dimension $3$, and a subalgebra given by $su(D)$. Here   $D=\dim(V^3) \dim(V^1)=4 \times 2=8$ acting on invariant spaces associated with the clusters $3$ and $4$. 
\eex

{\color{black}
\br{minimality}
In Theorem \ref{MainT}, every invariant subspace $S=F_1 \otimes F_2 \otimes \cdots \otimes F_N$ in (\ref{subspacesF}), is written in the form $E_1 \otimes E_2 \otimes \cdots \otimes E_{m_c}$ where each subspace $E_h={\large{\otimes}}_{j \in {\cal C}_h} F_j$ refers to one connected component of the associated graph. On this invariant subspace,  the action of the dynamical Lie algebra (and of the associated group of possible 
evolutions which is a subgroup of the   unitary group) is a tensor product action. Moreover, such a decomposition into invariant subspaces is {\it minimal} in the following sense: Given $E_1 \otimes E_2 \otimes \cdots \otimes E_{m_c}$ there is no other invariant subspace  $E_1^{'} \otimes E_2^{'} \otimes \cdots \otimes E_{m_c}^{'}$, with 
$E^{'}_h \subseteq E_h$,  $h=1,...,m_c$, where the strict inclusion holds for at least one $h$. This is due to the fact that every Lie algebra ${\cal G}_h$ in  (\ref{formaSDLA}) is an {\it  irreducible representation}, either of $su(2)$ or of $su(D^S_h) $ being the standard representation  for the given dimension $D^S_h$, which is also irreducible. 
\er

}

\section{Proof of Theorem \ref{MainT}} \label{prova}

\subsection{Casimir operators}

An important operator for what will follow will be the {\bf Casimir operator} 
$C^j$ (on the $j$-th space $(V^1)^{\otimes n_j}$ in (\ref{fullspace}))  defined as 
\be{Casimir}
C^j:=(S^j_x)^2+(S^j_y)^2+(S^j_y)^2, 
\ee
which is scalar on each irreducible representation $V^{f}$ with value on  $V^{f}$  given by $\frac{f}{2}(\frac{f}{2}+1)$ \cite{Woit}. {In particular, it is zero on (and only on) $V^0$.}  In the following,  operators will appear which are products of certain powers of the Casimir operator at certain locations in $\{1,...,N\}$ and other operators at other locations. For example $C^jS^k_x$, is the product of the Casimir operator at location $j$ with $S_x$ at location $k$, with $j\not=k$. Another example would be  $(C^j)^2C^lS^k_x$ with all different $j,k,l$ which is a square of $C^j$ together with $C^l$ and $S^k_x$. Another example is $A^j$ itself for an operator $A$ where all the powers of the Casimir operators are zero. Linear combinations of powers of Casimir operators form a (unital) commutative algebra. Therefore, their behavior in Lie brackets calculations when generating a given Lie algebra is easy  to control. We shall denote by $\Upsilon$ a general operator which is the product of Casimir operators. If we write 
$\Upsilon A^{j_1}B^{j_2} \cdots L^{j_r}$, we mean an operator which is $A$ in location $j_1$, $B$ in location $j_2$,...,$L$ in location $j_r$ and unspecified powers of Casimir operators in the remaining locations. If we want to point out the fact that these latest factors might be different from one operator to the other we use $\Upsilon_1 A^{j_1}B^{j_2} \cdots L^{j_r}$ and $\Upsilon_2 A^{j_1}B^{j_2} \cdots L^{j_r}$, for example.

\subsection{Reduction of the problem}

We first prove that  we can reduce ourselves to the following special case. 

\bp{Specialcase}
Assume that no subspace $F_j$ in (\ref{subspacesF}) is equal to 
$V^0$ and that the connectivity graph of the network is connected.  Then, if $N=1$, ${\cal G}_S$ is the representation of $su(2)$ associated with $F_1$. If $N \geq 2$ then ${\cal G}_S=su(D^S)$, with $D^S$ in (\ref{dimensioni}).  
\ep 
{Notice that if $F_j\neq V^0$ for all $j=1,\ldots,N$, then the connectivity graph of the network coincides with the associated graph relative to the invariant subspace.}

To see that the general case can be reduced to the special case  of Proposition \ref{Specialcase}, write the tensor product $S$ in (\ref{subspacesF}) by placing the $V^0$ spaces in the first $\bar N$ positions, i.e., like 
\be{firstspaces}
S=V^0 \otimes V^0 \otimes \cdots \otimes V^0 \otimes F_{\bar N+1} \otimes  F_{\bar N+2}\otimes \cdots \otimes F_{N},  
\ee
where $F_j=V^{r_j}$ with $r_j \geq 1$, for $j=\bar N+1,\ldots,N$. The dynamical 
Lie algebra ${\cal G}$ is generated by all the $S^j_{x,y,z}$ and by all the $S^j_z S^k_z$ for which the coupling constant $A_{j,k}$ are different from zero (Lemma \ref{Lemma1}). However,  on the subspace (\ref{firstspaces}) $S^j_{x,y,z}$,  $j=1,...,\bar N$ are all zero, { since, as we have mentioned when we introduced the Clebsch-Gordan decomposition,  $S_{x,y,z}^j$ is zero on the $V^0$ representation of $su(2)$.  For the same reason, $S^j_zS^k_z$, with $j<k$, and with   $j=1,...,\bar N$  are also zero. } Moreover zeros  so are also  all their (repeated) Lie brackets.  As a consequence, on these spaces,  the dynamical Lie algebra is the one generated by $iS^j_{x,y,z} $ and $iS^j_zS^k_z$, $j<k$, for all pairs $j$ and $k$ such that $A_{j,k} \not=0$ and  $j=\bar N+1,...,N$. 

The connectivity graph of the network of $N-\bar N$ clusters  of spins is not necessarily connected and coincides with the graph associated with the subspace (\ref{firstspaces}), i.e., the one obtained by removing the first $\bar N$ nodes and corresponding edges. Now by collecting in $ F_{\bar N+1} \otimes  F_{\bar N+2}\otimes \cdots \otimes F_{N}$ elements corresponding to the same connected components in order,  we notice that the element $S^j_zS^k_z$ and $S^j_{x,y,z}$  corresponding to pairs $(j,k)$ in the same connected component  generate a subalgebra which commutes with the ones correspoonding to the other connected components.  Therefore the whole subspace dynamical Lie algebra ${\cal G}_S$ takes the form  in  (\ref{formaSDLA}). 

Each term corresponds to one connected component of the associated graph and if we reduce ourselves to only one connected component the proof is reduced to the case of Proposition \ref{Specialcase}.

The case $N=1$ of Proposition \ref{Specialcase} follows immediately because if $N=1$ there is no interaction matrix of the form $S^j_zS_z^k$ but only the matrices $iS^1_{x,y,z}$ form the Lie algebra, which form indeed a representation of $su(2)$. The type of representation depends on the nature of the space $F_1$. 

The next subsections are devoted to prove the case $N \geq 2$ of  Proposition \ref{Specialcase}. 

\subsection{Generation of terms $S_{z}^j S_{z}^k$}

Lemma \ref{Lemma1} shows that the matrices $iS^j_zS^{k}_z$ belong to the dynamical Lie algebra ${\cal G}$ for every pair of clusters $j,k$ with nonzero coupling. The following Lemma shows that for a connected connectivity graph,  ${\cal G}$ contains matrices of the form $i \Upsilon S_{z}^j S_{z}^k$, for {\it any} pair of clusters $j, k$ (recall that $\Upsilon$ indicates a  general operator which is the product of Casimir operators)

{
\bl{connect}
Assume the connectivity graph of the network is connected. 
Then,  
for every pair $j < k \in \{1,2,...,N\}$, there exists in the dynamical Lie algebra ${\cal G}$ a matrix  
\be{newform}
i \Upsilon S_{z}^j S_{z}^k.
\ee
\el
}

\bpr
Fix two nodes $1\leq j<k\leq N$.
Given the connectedness  assumption of the graph, we know that there exists a path of length $r\geq 1$
of nodes $\hat n_i$, $i=0, \ldots, r$, with $\hat n_0=j$ and $\hat n_r=k$ such that $A_{\hat n_i,\hat n_{i+1}}\neq 0$.
 The claim will be proved by induction  on the length $r$ of the path joining the two nodes.

If $r=1$, the claim follows from Lemma \ref{Lemma1}. Assume $r>1$. Since the nodes $\hat n_0=j$ 
and $\hat n_{r-1}$ are connected by a path of length $r-1$, by 
the inductive assumption, we know that the  dynamical Lie algebra ${\cal G}$ contains a matrix of the type:
\be{uno-connect}
i\Upsilon
S^j_zS^{\hat n_{r-1}}_z,
\ee
Moreover since $A_{\hat n_{r-1},k}\neq 0$, we know from Lemma \ref{Lemma1}  that the matrix
\be{due-connect}
iS^{\hat n_{r-1}}_zS^k_z,
\ee
is  in  the dynamical Lie algebra ${\cal G}$ as well.
Since all the matrices of the type i$S^l_{x,y,z}$ are in ${\cal G}$, for any $l=1,...,N$, by taking  Lie brackets of the matrices in 
(\ref{uno-connect}) and (\ref{due-connect}), with these matrices we get that ${\cal G}$ contains 
all matrices of the type:
\be{tre-connect}
i\Upsilon
S^j_{x,y,z} S^{\hat n_{r-1}}_{x,y,z},
\ee
and 
\be{quattro-connect}
S^{\hat n_{r-1}}_{x,y,z}S^k_{x,y,z},
\ee
respectively. Notice that all $\Upsilon$ operators appearing in (\ref{tre-connect}) are the same. Now, we calculate (using (\ref{commurel2}))
\be{previous1}
\left[ i\Upsilon
S^j_{z} S^{\hat n_{r-1}}_{x}, iS^{\hat n_{r-1}}_{y}S^k_{z} \right ]=i \Upsilon 
S^j_{z}S^{\hat n_{r-1}}_{z}S^k_{z}, 
\ee
which belongs to ${\cal G}$ as well. Again,  since all the matrices of the type $i S^l_{x,y,z}$ are in ${\cal G}$, by taking Lie brackets of these matrices with the one in (\ref{previous1})  we get that:
\be{cinque-connect}
i \Upsilon S^j_{x,y,z}S^{\hat n_{r-1}}_{x,y,z}S^k_{x,y,z} \in {\cal{G}}, 
\ee
for all possible choices of $x,$ $y$ and $z$. 
Now,  we use matrices of type (\ref{tre-connect}) and (\ref{cinque-connect}), and we get:
\[
\left[  i \Upsilon S^j_{x} S^{\hat n_{r-1}}_{x},  i \Upsilon 
S^j_{y}S^{\hat n_{r-1}}_{x}S^k_{z} \right]=i \Upsilon_1 S^j_{z}(S^2_{x})^{\hat n_{r-1}}S^k_{z} \in {\cal G}
\]
By using $S^{\hat n_{r-1}}_{y,z}$ instead of $S^{\hat n_{r-1}}_x$ in the previous computation, we get that   the three matrices 
$$
i \Upsilon_1 S^j_{z}(S^2_{x,y,z})^{\hat n_{r-1}}S^k_{z}
$$
are all in ${\cal G}$, with the sames value for $\Upsilon_1$ for $x,y$, and $z$. By summing these matrices, using the definition of the Casimir operator (\ref{Casimir}),  we get
\[
i \Upsilon_2 S^j_{z} S^k_{z}\in {\cal G}, 
\] 
which is the claim of the Lemma. 

\epr

\subsection{Generation of terms $I_{zz}^j-I_{yy}^j$ and  $I_{yy}^j-I_{xx}^j$}

\bl{First} For every cluster $j=1,...,N$, there exists a matrix $ i \Upsilon (I^j_{zz}- I^j_{yy})$ and a matrix 
  $i \Upsilon  (I^j_{yy}- I^j_{xx})$ in  the dynamical Lie algebra ${\cal G}$.
\el
In the case where the $j$-th cluster contains only one spin $I^{j}_{(x,y,z)(x,y,z)}$ are taken equal to zero. So the statement is trivially true. 
\bpr Let us set $j=1$ (without loss of generality) and  $k=2$. 
We have that taking the Lie brackets between  
$i \Upsilon S_z^1 S^2_z$ (from Lemma \ref{connect})  and $S_{x,y}^1$ and  $S_{x,y}^2$, we obtain all possible 
$i \Upsilon  S_{x,y}^1 S^2_{x,y}$, and in fact, taking, possibly one extra Lie bracket with  $S_{x,y}^1$ or  $S_{x,y}^2$, we obtain all possible matrices 
\be{SxyzSxyz}
i \Upsilon S^1_{x,y,z}  S^2_{x,y,z} \in {\cal G}. 
\ee
Also observe from the calculation that the unspecified powers of Casimir operators in (\ref{SxyzSxyz}), which are collected in the term $\Upsilon$,  are the same for all the matrices in (\ref{SxyzSxyz}). Now consider 
\be{partialeq}
\left[ i \Upsilon  S_z^1 S_x^2, i \Upsilon  S_z^1  S_y^2 \right ]=i\Upsilon_1 {(S_z^1)^2}  S_z^2= i\Upsilon_1
({\frac{n_1}{4}{\bf 1}^1+2 I_{zz}^1}) S_z^2,
\ee
{\color{black} since, as it is easily seen by induction, on a space of $n_1$ spin $\frac{1}{2}$ of dimension $2^{n_1}$, 
 \be{basicrel}
 (S_g)^2=\frac{n_11}{4}{\bf 1}+2 I_{gg}, \qquad \text{for} \qquad  g=x,y,z.
 \ee 
 Now, {by using $S^1_y$ instead of $S^1_z$ in (\ref{partialeq}) we obtain the matrix $i\Upsilon_1 {(S_y^1)^2}  S_z^2= i\Upsilon_1 ({\frac{n_1}{4}{\bf 1}^1+2 I_{yy}^1}) S_z^2$. Taking the difference between this matrix  and the one in (\ref{partialeq})
 we obtain that $i \Upsilon_2 ( I_{zz}^1- I_{yy}^1)(S_z^2)^2$ belongs to ${\cal G}$}.}
 With analogous calculations, replacing $S_z^2$ with $S_x^2$ or $S_y^2$ we obtain also $i \Upsilon_2 (I_{zz}^1-I_{yy}^1) (S_x^2)^2$ and $i \Upsilon_2 (I_{zz}^1-I_{yy}^1) (S_y^2)^2$. 
 It is important to notice at this point that since the omitted Casimir operators in (\ref{SxyzSxyz}) are all equal and the sequence of calculation is the same in all three cases (with $x,y,$ or $z$ on the right hand side), the omitted Casimir operators (in the operator $\Upsilon_2$)  are the same in all three cases. We can therefore sum these three matrices and obtain using the  definition of the Casimir operator (\ref{Casimir}) that $i \Upsilon_3 (  I_{zz}^1- I_{yy}^1)$ belongs to ${\cal G}$, for some $\Upsilon_3$ operator. A completely 
 analogous calculation gives that  $i \Upsilon ({ I^1}_{yy}- I^1_{xx})$ also belongs to ${\cal G}$, 
 for some $\Upsilon$ operator. 
\epr

\subsection{Lie subalgebra of $u(2^n)$ commuting with the symmetric group}

We now  need to recall some general facts on the Lie  subalgebra of $u(2^n)$ of matrices commuting with the permutation group $P_n$. Denote this subalgebra as $u^{P_n}(2^n)$. Its dimension is given by (cf. \cite{NoiJMP}) 
\be{dimen}
\dim\left( u^{P_n}(2^n) \right)={{n+3}\choose{n}}.
\ee
One of the main results of \cite{NoiJMP} is the following 
\bt{fromJMP}
$\{iI_{zz}, iS_{x,y,z} , i{\bf 1} \}$ generate $u^{P_n}(2^n)$, and $\{iI_{zz}, iS_{x,y,z} \}$ generate 
$u^{P_n}(2^n)\cap su(2^n)$.
\et
As we already recalled, the space $(V^1)^{\otimes n}$ decomposes  according to (iterated) 
Clebsch-Gordan decomposition of a tensor product representation in the direct sum of (possibly repeated) 
$V^n$, $V^{n-2}$,..., irreducible representations of $su(2)$. Since $S_{x,y,z}$ and $I_{zz}$ leave such subspaces invariant,\footnote{To see this for $I_{zz}$ recall that $S_z^2=\frac{n}{4}{\bf 1} +2 I_{zz}$ (from (\ref{basicrel})) so that $I_{zz}=\frac{1}{2}(S_z^2-\frac{n}{4} {\bf 1} )$.} these spaces are invariant for $u^{P_n}(2^n)$ as well, because of Theorem \ref{fromJMP}.  Therefore, in coordinates given by the bases of these spaces,  the matrices of 
$u^{P_n}(2^n)$ take a block diagonal form.\footnote{In \cite{ConJonas} such a block diagonal form was described using a different approach based on Young symmetrizers.}  Consider two subspaces in the 
 decomposition of the form $V^f$ for some $f$, i.e., two subspaces of the same dimension, say $V^f_1$ and $V^f_2$. A basis for these spaces can be obtained starting with the {\it highest weight vector}  and then successively applying the lowering operator as described for example in \cite{Woit}. The operators 
$S_{x,y,z}$ and therefore $I_{zz}=\frac{1}{2}(S_z^2-\frac{n}{4} {\bf 1})$  as well as the identity $i{\bf 1}$  act in the same way on these bases, and therefore (by induction), each repeated  Lie bracket of them. Therefore we can take a basis so that the blocks of $u^{P_n}(2^n)$ of the same 
dimension are equal to each other. Furthermore, each block of dimension $f+1$ can take  
any value in $u(f+1)$ independently of the other blocks of different dimensions, that is, for each block of dimension $f+1$ there are $(f+1)^2$ degrees of freedom. If this  was not the case for one block, 
we would have a total number of degrees of freedom,  {\it which is the dimension of $ u^{P_n}(2^n)$},   strictly less than $T_n$, where $T_n$ is defined, for $n$ odd, as 
\be{Tnodd}
T_n=2^2+4^2+\cdots + (n+1)^2, 
\ee
 and, for $n$ even,  as 
 \be{Tneven}
 T_n=1^2+3^2+\cdots + (n+1)^2. 
 \ee
However in both cases, $n$ odd in (\ref{Tnodd}) and $n$ even in (\ref{Tneven}),  an induction argument shows that 
$$
T_n={{n+3}\choose{n}}, 
$$ 
which is from (\ref{dimen}) the dimension of $ u^{P_n}(2^n)$. So we obtain a contradiction. Therefore, we have the following form of Theorem \ref{fromJMP}, which will be useful for us 
\bc{fromJMP2} The restrictions of $\{iI_{zz}, iS_{x,y,z},  i{\bf 1} \}$ to every irreducible representation $V^f$ of $su(2)$ generate $u(f+1)$.
\ec

\subsection{Controllability on a single factor in (\ref{subspacesF})}

We now show a notion of controllability on each factor $F_j$ in (\ref{subspacesF}). Recall that each of these factors is assumed of the form $V^f$, with $f\geq 1$ in Proposition \ref{Specialcase}, although the next lemma can be stated without restrictions on $f$.

\bl{Oneextrastep} Fix any $j \in \{1,...,N\}$ with $F_j$ in (\ref{subspacesF}) equal to $F_j=V^f$ so that 
 $f+1=\dim(F_j)$. Then for every 
$M \in su(f+1)$ the dynamical Lie algebra ${\cal G}$ contains a matrix $i\Upsilon  A^j$ such that 
the restriction of $iA^j$ to $F_j$ is equal to $M$. 
\el

\bpr
As we have done above, to simplify notations, we set, without loss of generality $j=1$. The statement 
is trivially true (and not useful for us because we are assuming in Proposition \ref{Specialcase} that all $F_j$ have dimensions strictly larger than $1$)  if $\dim(F_1)=1$ and it is  also true in the case $\dim(F_1)=2$ since $iS^j_{x,y,z}$ belong to ${\cal G}$. 

It is useful to use the notation $\langle B_1,...,B_s \rangle$ for the Lie algebra generated by certain matrices 
$\{B_1,...,B_s \}$ so that, for instance,  the first statement of Theorem \ref{fromJMP} reads as 
$\langle iI_{zz}, iS_{x,y,z}, i{\bf 1} \rangle=u^{P_n}(2^n)$. Denote by $n_1$ the number of spin $\frac{1}{2}$ particles in the first cluster. Consider the matrix 
$Q^1:=I^1_{xx}+I^1_{yy}+I^1_{zz}=\frac{1}{2}(C^1-3\frac{n_1}{4}{\bf 1}^1)$, with the Casimir operator (\ref{Casimir}) on the first set, which commutes with every 
matrix in $\{i( I^1_{zz}-I_{yy}^1), i( I^1_{yy}-I_{xx}^1), iS_{x,y,z}^1 \}$ (and therefore with each repeated Lie bracket of them). Then we have by Theorem \ref{fromJMP}
\be{basicCompu}
\left( su(2^{n_1}) \cap u^{P_{n_1}}(2^{n_1}) \right) \otimes{\bf 1} \otimes {\bf 1} \otimes \cdots \otimes {\bf 1} = 
\langle iI_{zz}^1, iS_{x, y, z}^1 \rangle \subseteq \langle i( I^1_{zz}-I_{yy}^1), i( I^1_{yy}-I_{xx}^1), iS_{x,y, z}^1, iQ^1 \rangle = 
\ee
$$
\langle  i(I^1_{zz}-I_{yy}^1), i( I^1_{yy}-I_{xx}^1), i S_{x,y, z}^1 \rangle + \texttt{span}(i Q^1) \subseteq u^{P_{n_1}}(2^{n_1}) \otimes {\bf 1} \otimes {\bf 1} \otimes \cdots \otimes {\bf 1}. 
$$
In the first equality, we used Theorem \ref{fromJMP} and in the second equality we used the commutativity of $Q^1$. Now, consider relation (\ref{basicCompu}) in the basis where matrices are 
 block diagonal and in particular on the block corresponding to 
 $F_1 \otimes F_2 \otimes \cdots \otimes F_N$ in (\ref{subspacesF}). Restricting to this block we notice that $\texttt{span}(Q^1)$ is included in the span of the identity on it (it commutes with an irreducible representation of $su(2)$ given by the restriction of $\texttt{span} \{ iS_x^1, i  S_y^1,i S_z^1)  \}$  and therefore it must be a multiple of the identity according to Schur's lemma (see, e.g., \cite{Woit})). Consider now the block diagonal form of the relation (\ref{basicCompu}), and its form on the block corresponding to $F_1 \otimes F_2 \otimes \cdots \otimes F_N$. 
 The first Lie algebra  on the left  is $su(f+1) \otimes {\bf 1} \otimes {\bf 1} \otimes \cdots \otimes {\bf 1}$,   the second to last Lie algebra   is the restriction of $\langle i( I^1_{zz}-I_{yy}^1), i( I^1_{yy}-I_{xx}^1), i(S_{x,y,z}^1) \rangle$ to $F_1$  plus the span of the identity, everything tensored by the identity $N-1$ times. The last Lie algebra is  $u(f+1) \otimes {\bf 1} \otimes {\bf 1} \otimes \cdots \otimes {\bf 1}$. Now, using the fact from Lemma 
 \ref{First} that ${\cal G}$ contains  $\{i\Upsilon(I^1_{zz}- I_{yy}^1), i \Upsilon(  I^1_{yy}- I_{xx}^1), 
i  S_{x,y,z}^1 \}$ and that Casimir operators are {\it all non zero} on the subspaces $F_2,F_3,...,F_N$ because of our assumption on the dimension, it follows that we can generate every element of the restriction of 
  $\langle i( I^1_{zz}-I_{yy}^1), i( I^1_{yy}-I_{xx}^1), iS_x^1, i S_y^1,i S_z^1 \rangle=su(2^{n_1}) \cap u^{P_{n_1}}(2^{n_1})$ to $F_1$. This concludes the proof.

\epr

 \subsection{Maximal subalgebras in $su(rs)$} 
 
 Now that we know that ${\cal G}$ acts as any desired  element of $su(f+1)$ on any factor in (\ref{subspacesF}) we need to 
 show that from these  elements we can generate all of $su(D^S)$ with $D^S$ in (\ref{dimensioni}). Recall that  ${\cal G}$ also contains $i \Upsilon  S_z^j S_z^k$ for every pair $j,k$ according to Lemma \ref{connect}.  Denote by 
$f_j+1$, $j=1,...,N$ the dimension of $F_j$. According to Lemma \ref{Oneextrastep} we have on 
$F_1 \otimes F_2 \otimes \cdots \otimes F_N$, $su(f_1+1) \otimes {\bf 1} \otimes {\bf 1} \otimes \cdots \otimes {\bf 1}$,  $ {\bf 1} \otimes su(f_2+1)  \otimes {\bf 1} \otimes \cdots \otimes {\bf 1}$,..., $ {\bf 1}   \otimes {\bf 1} \otimes \cdots \otimes {\bf 1} \otimes su(f_N+1)$, besides the restriction of $i \Upsilon  S_z^j S_z^k$. We will apply iteratively the following result 
\bt{fromDynkin}
For each pair $r,s \geq 2$, the Lie algebra which is a direct sum of $su(r) \otimes {\bf 1}$ and ${\bf 1} \otimes su(s)$ is a maximal Lie algebra of $su(rs)$. 
\et 
A maximal Lie algebra ${\cal L} \subseteq su(rs)$ is by definition such that for every element 
$A \in su(rs)$ with $A \notin {\cal L}$, $\langle A, {\cal L} \rangle =su(rs)$.  Theorem \ref{fromDynkin} 
 was proved by E.B. Dynkin in \cite{Dynkinpaper} (Theorem 1.3 in that paper). We only need a simpler version of it,  which says that for each $A \otimes B $, $\notin su(r) \otimes {\bf 1}$ and $\notin {\bf 1} \otimes su(s)$, 
$\langle A\otimes B, su(r) \otimes {\bf 1},  {\bf 1} \otimes su(s)\rangle\rangle =su(rs)$. In order to see this,  consider 
$$
{+}_{m=0}^\infty ad_{{\bf 1} \otimes su(s)}^m A \otimes B=
A\otimes \left(  {+}_{m=0}^\infty ad_{su(s)}^m B \right). 
$$  
Since ${+}_{m=0}^\infty ad_{su(s)}^m B$ is a nonzero ideal in $su(s)$ and $su(s)$ is simple, it must be equal to $su(s)$. Therefore for every matrix $C \in su(s)$ we have that $A \otimes C$ belongs  to the generated 
Lie algebra. Fixing $C$,  and doing the same thing on the left, we have that for every $E \in su(r)$, $E\otimes C$ also belongs to the generated Lie algebra. Therefore, in conclusion,  such a Lie algebra contains all the matrices of the form $E \otimes C$ with $E \in su(r)$ and $C \in su(s)$ beside $su(r) \otimes {\bf 1}$ and ${\bf 1} \otimes su(s)$. Putting these together, they  span of  $su(rs)$.

\subsection{Conclusion of the proof}

The proof of the Proposition \ref{Specialcase} and therefore of the theorem is  completed as follows. On the space $F_1 \otimes F_2$, we have $su(f_1+1) \otimes {\bf 1}$ and $ {\bf 1} \otimes su(f_2+1) $  along with the restriction of 
$i\Upsilon S_z^1  S_z^2$ which is nonzero because all the restriction of all the Casimir operators are nonzero multiples of the identity, and it is not in $su(f_1+1) \otimes {\bf 1}$ nor in $ {\bf 1} \otimes su(f_2+1) $. Therefore, using Theorem \ref{fromDynkin}, we have that ${\cal G}$ contains matrices that are equal to $M$ for any $M \in su((f_1+1)(f_2+1))$ on $F_1 \otimes F_2$ and equal to the identity on the other factors in (\ref{subspacesF}). Then we iterate this argument by using $i \Upsilon  S_z^2  S_z^3$ to show this fact for $M \in su((f_1+1)(f_2+1)(f_3+1))$, $i \Upsilon S_z^3 S_z^4$ and so on up to $i \Upsilon S_z^{N-1}  S_z^N$ for $M \in su(D^S)$.

\vspace{0.25cm}

{

\section{Discussion and  Extensions}\label{Exte}

We now discuss  several possible extensions of the result of Theorem \ref{MainT} to networks  different from the {\it multipartite} case   with Ising coupling above treated.

\subsection{Networks with different couplings between spins}

The Ising coupling between spins in two different clusters, $A_{j,k} S_z^jS_z^k$, can be replaced by a more general two body coupling so that $A$ in (\ref{A}) is replaced by $\hat A$ with 
\be{newinter}
i\hat A=\sum_{1\leq j < k \leq N} A_{j,k}S_z^jS_z^k+B_{j,k} S_x^jS_x^k+C_{j,k} S_y^jS_y^k. 
\ee
The result of Theorem \ref{MainT} is still valid as long as we consider as a non-zero interaction
 between the $j$-th and the $k$-th cluster if $(A_{j,k}, B_{j,k}, C_{j,k}) \not= (0,0,0)$. In order to see this, notice that the subspaces (\ref{subspacesF}) are still invariant for the dynamics if the interaction takes the more general form (\ref{newinter}) and that the reduction to the case of Proposition \ref{Specialcase} still holds. If there is only one cluster in the network, there is no term of the two-body form (\ref{newinter}) and so the result of the proposition holds. If there is more than one cluster in a connected network we have proven subspace controllability in the Ising $Z-Z$ case. Let us see why this is true in the general case of interaction     (\ref{newinter}). By taking repeated Lie brackets of the interaction (\ref{newinter}) with matrices of the form $iS_{x,y,z}^j$ we can obtain (as long as the coupling is nonzero) the Ising terms  $iS_z^j S_z^k$. Therefore, the dynamical Lie algebra generated by replacing the Ising interaction (\ref{A}) with the more general (\ref{newinter}) two body interaction is larger than or equal to the one obtained with Ising interaction. Since in the latter case we have subspace controllability,  the same is true  for the more general interaction (\ref{newinter}). 
 
 \subsection{Coupling between spins in the same cluster}
 
If we add to the interaction $A$ in (\ref{A}) a term modeling interaction between spins in the same cluster, the coupling takes the more general form 
\be{iAgen}
iA_{gen} =iA+\sum_{j=1}^NH_0^j, 
\ee 
 where $A$ is the same as in (\ref{A}) (or (\ref{newinter})) and $H_0^j$ models these 
 `internal' interactions. By using  the form of the interaction $A$ in (\ref{A}) and taking repeated Lie brackets of (\ref{iAgen}) with 
 $S_{x,y,z}^j$, $j=1,...,N$, we can detach $iA$ from $iA_{gen}$ in (\ref{iAgen}) and therefore the dynamical Lie algebra in this case is generated by the same dynamical Lie algebra calculated above for the case  without internal interactions, and the matrix $\sum_{j=1}^NiH_0^j$ (which can also be separated into pieces with the same technique). Therefore the dynamical Lie algebra will be in general larger and the spaces (\ref{subspacesF}) will in general not be invariant anymore.

\subsection{Different couplings for  spins in the same cluster}

As it is intuitive, if we allow spins of the same cluster to interact differently with the same spin in another cluster we increase the controllability of the system in that some of the subspaces (\ref{subspacesF}) will not be invariant anymore and larger invariant subspaces have to be considered. We illustrate this fact  with a simple  example.

\bex{Exaplus1}
Consider first two spin $\frac{1}{2}$ particles with the same gyromagnetic ratio 
interacting in the same way with one spin $\frac{1}{2}$ particle with a different gyromagnetic ratio. We have two clusters with two and one spin respectively. On the first cluster, the Hilbert space $V^1 \otimes V^1$ splits according to Clebsch-Gordan decomposition as $V^1 \otimes V^1=V^2 \oplus V^0$ so that the full space $(V^1 \otimes V^1) \otimes V^1$ splits as $(V^2 \otimes V^1) \oplus  (V^0 \otimes V^1)$. Therefore the spaces $V^2 \otimes V^1$ and $V^0 \otimes V^1\simeq V^1$ are the ones to be considered in (\ref{subspacesF}). In the first case the associated graph coincides with the connectivity graph of the network, the dimension $D^S$ in (\ref{dimensioni}) is equal to $D^S=6$,  and the dynamical Lie algebra acting on this invariant space coincides with $su(6)$. In the second case the associated graph only has the node corresponding to the second cluster. The dynamical Lie algebra on the given subspace coincides with $su(2)$ (its irreducible standard representation). Therefore in the appropriate basis, the dynamical full Lie algebra ${\cal G}$ can be written in block diagonal form, with blocks of dimension $6$ and $2$. However, if the coupling constants are different in absolute value, a direct calculation of the dynamical Lie algebra shows that it is equal to $su(8)$. Therefore, there is no nontrivial  invariant   subspace and the system is controllable as a whole. The two subspaces above are included in a single invariant subspace equal to the whole space. 
\eex

We now want to obtain some insight into  the mechanism of increase in controllability and 
enlargement of the invariant subspaces when the coupling constants differ which we have seen in the previous example.  We start with the 
 basic situationof  the type of networks considered in the previous sections and then perturb some coupling constants. Consider, in particular, a network with $N$ clusters as in the previous sections, each cluster with uniform coupling with any other cluster. Consider then an associate invariant subspace as in (\ref{subspacesF}). Assume now that the coupling constants of one of the cluster, say the cluster $N-1$,  with another cluster,  say the $N$-th cluster, {split}.  A subcluster of the $(N-1)$-th cluster has coupling constant with the $N$-th cluster equal to $W$ and another subcluster has coupling constant $Y$ (assuming for simplicity that there are only two values of coupling constants and furthermore assume the stronger condition $|Y|\not=|W|$). The matrix $A$
 in (\ref{A}) can then  be written as 
\be{newA}
iA:=\sum_{1\leq j < k \leq N, \,  (j,k)\not= (N-1,N)} A_{j,k}S_z^jS_z^k + WS_{z,1}^{N-1}S_z^N+YS_{z,2}^{N-1} S_z^N, 
\ee
   where we have split $S_z^{N-1}$ in two parts,  $S_{z,1}^{N-1}$ and $S_{z,2}^{N-1}$, according to their interaction with the $N$-th cluster. Now, if $F_N=V^0$, the last two terms in (\ref{newA}) as well  as all the coupling $S^j_zS^N_z$ and also  $S_{x,y,z}^N$ give zero, the associated graph to the subspace (\ref{subspacesF}) only contains the first $N-1$ nodes. The splitting of the coupling constants in the cluster $N-1$ plays no role and the situation is equivalent to the one 
   we considered in the previous sections but with the first $N-1$ clusters only. If however,  $F_N\not=V^0$,  by  taking (repeated) Lie brackets of $A$ in (\ref{newA}) with $iS^N_{x,y,z}$ in and $iS_{x,y,z}^{N-1}$ we obtain all matrices of the form 
$
i WS_{x,y,z,1}^{N-1} S_{x,y,z}^N
+ i YS_{x,y,z,2}^{N-1} S_{x,y,z}^N
$
where we have split $S_{x,y,z}^{N-1}$ as $S_{x,y,z}^{N-1}=S_{x,y,z,1}^{N-1}+S_{x,y,z,2}^{N-1}$, generalizing what we have done above. Taking the Lie brackets of 
$
i WS_{x,1}^{N-1} S_{x}^N
+ i YS_{x,2}^{N-1} S_{x}^N
$
with $
i WS_{y,1}^{N-1} S_{x}^N
+ i YS_{y,2}^{N-1} S_{x}^N
$, 
we obtain $
\left(i W^2S_{z,1}^{N-1}
+ i Y^2S_{z,2}^{N-1} \right) (S_{x}^N)^2
$. 
 Analogously we obtain $
\left(i W^2S_{z,1}^{N-1}
+ i Y^2S_{z,2}^{N-1} \right) (S_{y}^N)^2
$ and $
\left(i W^2S_{z,1}^{N-1}
+ i Y^2S_{z,2}^{N-1} \right) (S_{z}^N)^2
$, 
and summing all we obtain 
\be{obtain3}
\left(i W^2S_{z,1}^{N-1}
+ i Y^2S_{z,2}^{N-1} \right) C^N, 
\ee
where $C^N$ is the Casimir operator. Analogously, we can obtain (\ref{obtain3}) with $z$ replaced  by $x$ and $y$, respectively. Since $C^N$ is a multiple of the identity on $F_N$, we effectively obtain $W^2 S_{x,y,z,1}+Y^2S_{x,y,z,2}$ and since we already had $S_{x,y,z}=S_{x,y,z,1}+S_{x,y,z,2}$ we obtain the two matrices $S_{x,y,z,1}$ and $S_{x,y,z,2}$. We have effectively split the cluster $N-1$ into two subclusters. The subspace $F_{N-1}$ is not invariant anymore.
If we reconsider the separation of the $(N-1)-$th cluster into the two subclusters as above we can apply the Clebsch-Gordan decomposition to each subcluster. {Assume that the first subcluster has $m_1$ spins and the second $m_2$ 
(thus $n_{N-1}=m_1+m_2$),  we will have a decomposition of $(V^1)^{\otimes m_1}$ for the first subcluster and  a decomposition of $(V^1)^{\otimes m_2}$  for the second subcluster. }
Pick a space in the first decomposition, say $V^{f_1}$ and a space in the second decomposition, say $V^{f_2}$, which carry respectively an irreducible representation corresponding to $f_1$ and $f_2$ of $su(2)$. To $V^{f_1} \otimes V^{f_2}$ we can apply the Clebsch-Gordan decomposition into the direct sum of invariant subspaces. The original invariant subspace $F_{N-1}$ was selected among such spaces. However, with the division into two subclusters above, the tensor product $V^{f_1} \otimes V^{f_2}$ has to be considered as a whole, giving therefore a larger invariant space.

\section*{Acknowledgement} D. D'Alessandro research was  supported by   the NSF 
under Grant ECCS 1710558.

\end{document}